# Magnetoelectric Effect in Ni-PZT-Ni Cylindrical Layered Composite Synthesized by Electro-deposition

D. A. Pan, Y. Bai, W. Y. Chu and L. J. Qiao

Environmental Fracture Laboratory of Education of Ministry, Corrosion and Protection Center, University of Science and Technology Beijing, Beijing 100083, P. R. China

E-mai:  lqiao@ustb.edu.cn

**Abstract**: The magnetoelectric (ME) coupling of cylindrical trilayered composite was studied in this paper. The Ni-lead zirconate titanate (PZT)-Ni trilayered cylindrical composite was synthesized by electro-deposition. The maximum ME voltage coefficient of cylindrical ME composite is 35V/cm Oe, about three times higher than that of the plate trilayered composite with the same raw materials and magnetostrictive- piezoelectric phase thickness ratio. The high ME voltage coefficient of cylindrical composite owes to the self-bound effect of circle. Moreover, the resulting complex condition can induce a double peak in the field dependence of ME coefficient.

## 1. Introduction

Multiferroic materials have drawn increasing interest due to their multi-functionality, which provides significant potential for applications in the next-generation multifunctional devices [1]. In the multiferroic materials, the coupling interaction between ferromagnetic and ferroelectric can produce some new effects, such as magnetoelectric (ME) effect and magnetodielectric effect [2].

From 1970s, many particulate [3-5] and in-situ-grown [6-7] magnetoelectric composites have been developed by using piezoelectric materials and magnetostrictive ferrites to overcome the problems of single-phase magnetoelectric materials, i.e. low magnetoelectric response and requirement of low temperature [8]. However, these composites have shortcomings in reproducibility and reliability, such as control of the connectedness of phases, and their magnetoelectric voltage coefficient is also insufficient for practical applications [5].

To obtain higher ME coefficient, various layered ME composites, such as $Tb_{1-x}Dy_xFe_{2-y}$ (Terfenol-D)/ Pb(ZrTi)$O_3$ (PZT), ferrite / PZT, and Terfenol-D/PVDF laminates, were widely investigated in the past several years, where giant magnetostrictive materials and giant piezoelectric materials were selected as the components [9-15]. In these layered ME composites, the ME coefficient depends on the magnetic-mechanical-electric coupling between magnetostrictive layer and piezoelectric layer, so interfacial binding also greatly influences the ME coefficient of layered composites. To improve the interfacial binding between different layers, we developed ME layered composite by electro-deposition. A high ME voltage coefficient of 13V/cm Oe was obtained in a Ni/PZT/Ni trilayered composite. Moreover, electro-deposition can be used to fabricate magnetoelectric coupling devices with complex shapes and easily control the thickness of each layer. That overcomes the limitation of previous preparation method, that only simple shape layer, such as disk, square and rectangle, can be made into ME laminates. Wan [16] reported that the ME voltage coefficient of the arched trilayered composite was larger than that of the plate trilayered composite. It indicates that the ME voltage coefficient is able to be enhanced through improving the shape or configuration of ME composite. In this work, cylindrical Ni-PZT-Ni trilayered composites were synthesized by electro-deposition, and high magnetoelectric coefficient was obtained.

**2. Experimental details**

After materized and jointed electrodes on outer and inner sides, A PZT cylinder (20mm outer diameter, 18mm inner diameter and 8mm height) was polarized at 425 K in an electric field of 30-50 kV/cm along radial direction.(Fig. 1) Then, PZT cylinder was bathed in nickel aminosulfonate plating solution, and was electrodeposited Ni on both sides. After 10 hours of electro-deposition, the total thickness of Ni is about 1mm.

During ME measurement, samples were subjected to a bias magnetic field $H_{DC}$ superimposed with an AC field $\delta H$ (1 kHz-120 kHz). The generated voltage $\delta V$ across the thickness of the cylinder was amplified and measured by an oscilloscope. Since AC magnetic field $\delta H$ was generated by a Helmholtz coil, the amplitude of AC magnetic field $\delta H$ = 22 Oe when the amplitude of AC current is equal to 1A through the coil. The ME voltage coefficient was calculated according to $\alpha_E = \delta V /(t_{PZT} \cdot \delta H)$, where $t_{PZT}$ is the thickness of PZT. In the experiment, two ME voltage coefficients of $\alpha_{E,A}$ and $\alpha_{E,V}$ were obtained corresponding to two conditions, $H_{DC}$ and $\delta H$ along or vertical to the axis of cylinder.(Fig. 1)

**3. Results and discussion**

Firstly, the dependence of $\alpha_{E,A}$ and $\alpha_{E,V}$ on bias magnetic field $H_{DC}$ was measured at 1 kHz.(Fig.2) With the enhancement of axial magnetic field, $\alpha_{E,A}$ increases first, reaches a maximum at $H_m$=450 Oe, and then decreases rapidly. In contrast, with the enhancement of vertical magnetic field, except a shape peak of $\alpha_{E,V}$ at $H_{m1}$=160 Oe, a second flat peak appears at $H_{m2}$=4100 Oe.

For a plate layered composite, as $H_{DC}$ and $\delta H$ are applied along the length direction, a sharp peak of $\alpha_E$ always appears at the low field. However, as $H_{DC}$ and $\delta H$ are applied along the thickness direction, a flat peak of $\alpha_E$ appears at the high field.(Fig. 3a-b) That difference originates from the influence of shape demagnetization on the magnetostriction of the magnetostrictive phase[17]. The ME coefficients are directly proportional to magnetostrictive coefficient $q \sim \delta \lambda/\delta H$, where $\delta\lambda$ is the magnetostriction. The plate shape of magnetostrictive phase determines that in-plane magnetostriction will be much easier, so the maximum of $\alpha_E$ is larger and the corresponding frequency is lower.

The cylindrical ME composite can be divided to a series of infinitesimal plate units. (Fig. 1) In the axial mode, the fields are applied along the length direction of each infinitesimal plate unit, so a sharp peak of $\alpha_{E,A}$ appears at low field. As the magnetic fields are vertical to the axis of cylinder, the condition will be more complex. In this condition, each infinitesimal unit slopes in the fields, and can be regarded as the parallel combination of two plate units, one parallel to the fields and the other normal to the fields. As a result, a sharp peak at low field and a flat peak at high field appear simultaneously in the field dependence of $\alpha_{E,V}$. Moreover, a plate ME composite sloping in the field and the parallel combination of two plate ME composites were measured.(Fig.3 c-d) Similar curves with double peaks in the field dependence of ME voltage coefficient confirm our above statement on cylindrical ME composite.

Figure 4 shows the frequency dependence of $\alpha_{E,A}$ and $\alpha_{E,V}$ measured at a fixed bias field. For both $\alpha_{E,A}$ and $\alpha_{E,V}$, there is a sharp resonance peak at $f \approx 63.8$ kHz, where the large ME coefficient is associated with the electromechanical resonance (EMR)[18]. Similar resonance was also reported in plate trilayered composites [19]. For the cylindrical ME composite with the thickness ratio of $t_{Ni}/(t_{Ni}+t_{PZT})=1/2$, the maximum ME voltage coefficient is $\alpha_{E,V}=35$ V/cm Oe at 160 Oe, which is about three times larger than that of a plate trilayered composite with the same magnetostrictive-piezoelectric phase ratio, $\alpha_{E,31}=13$V/cm Oe.(Fig. 5)

The higher ME coefficient of cylindrical ME composite originates from the complex stress condition due to the self-bound effect of circle. For a plate composite in free boundary condition, only one piezoelectric mode of $d_{31}$ in piezoelectric phase contributes to ME coupling, while the much higher $d_{33}$ does not play a role. But the stress condition in a cylindrical ME composite is much more complex. When the magnetostrictive Ni ring expands in the magnetic fields, not only the circumference increases, but also the diameter rises at the same time due to the self-bound effect of circle. Then each piezoelectric PZT infinitesimal unit will suffer a radial force and a tangent force simultaneously due to the shape change of Ni layers. Two piezoelectric modes of $d_{33}$ and $d_{31}$ in PZT contribute to the ME coefficient at the same time, so the cylindrical ME composite has a higher ME voltage coefficient than plate one. It was reported that the clamped plate ME composite had larger ME voltage coefficient than that in free boundary condition, because both $d_{31}$ and $d_{33}$ worked for the ME coupling [20]. Although in our experiment not an external mechanical force but the self-bound effect of circle changes the free boundary condition of ME composite, it is similar that $d_{33}$ plays an important role in the high ME coefficient.

## 4. Summary

In summary, the Ni-PZT-Ni trilayered cylindrical composite was synthesized by electro-deposition. The maximum ME coefficient of the cylindrical trilayered composite is 35 V/cm Oe, which is about five times higher than that of the plate trilayered composite with the same magnetostrictive- piezoelectric phase thickness ratio. Electro-deposition provides an efficient method to synthesize various ME composites with complex structures, so that better ME properties can be obtained though structure design. The improvement of preparation method may promote the development of ME composite towards practical application notably.

### Acknowledgment
This project was supported by program for Changjiang Scholars, Innovative Research Team in University (IRT 0509) and the National Natural Science Foundation of China under Grant No. 50572006.

# Figure Captions Page

FIG. 1. Structure sketch of cylindrical layered composite.

FIG. 2. Dependence of ME voltage coefficient $\alpha_{E,A}$ (a) and $\alpha_{E,V}$ (b) on bias field $H_{DC}$ at $f$=1 kHz of AC field $\delta H$ for the Ni-PZT-Ni cylindrical layered composite.

FIG. 3. Dependence of ME voltage coefficient on field $H_{DC}$ at $f$=1 kHz of Ni-PZT-Ni plate composite with dimension of 10×20×0.8 mm$^3$; (a) the plate sample parallel to the field, (b) the plate sample vertical to the field, (c) the parallel connection of two plate samples, (d) the plate composite sloping in the field, $\theta$=45º.

FIG. 4. Frequency dependence of ME voltage coefficient of Ni-PZT-Ni cylindrical layered composite: (a) $\alpha_{E,A}$ at $H_m$=450 Oe, (b) $\alpha_{E,V}$ at $H_{m1}$=160 Oe, (c) $\alpha_{E,V}$ at $H_{m2}$=4100 Oe.

FIG. 5. Frequency dependence of ME voltage coefficient $\alpha_{E,31}$ of Ni-PZT-Ni plate layered composite with dimension of 10×20×0.8 mm$^3$ at $H_{DC}$=$H_m$. The total thickness of Ni is about 0.8mm.

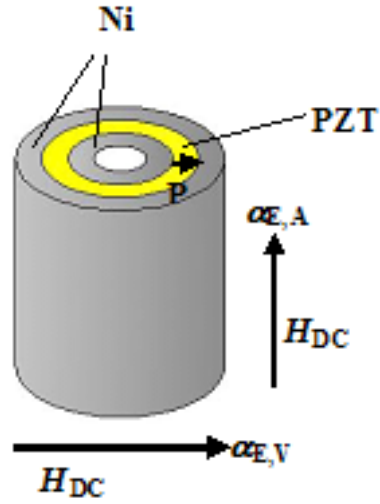

FIG. 1. Structure sketch of cylindrical layered composite.

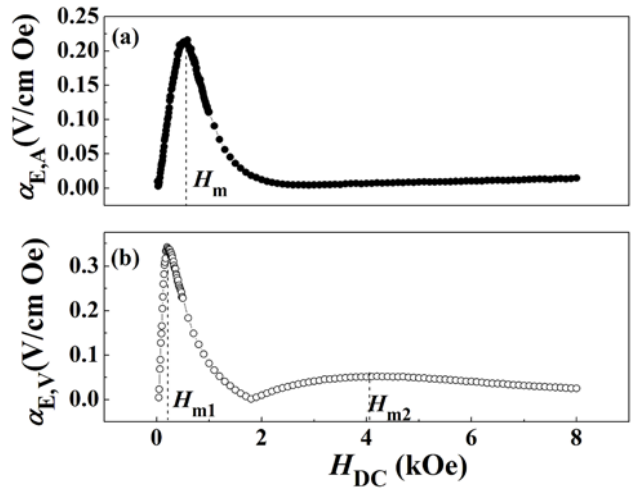

FIG. 2. Dependence of ME voltage coefficient $\alpha_{E,A}$ (a) and $\alpha_{E,V}$ (b) on bias field $H_{DC}$ at $f$=1 kHz of AC field $\delta H$ for the Ni-PZT-Ni cylindrical layered composite.

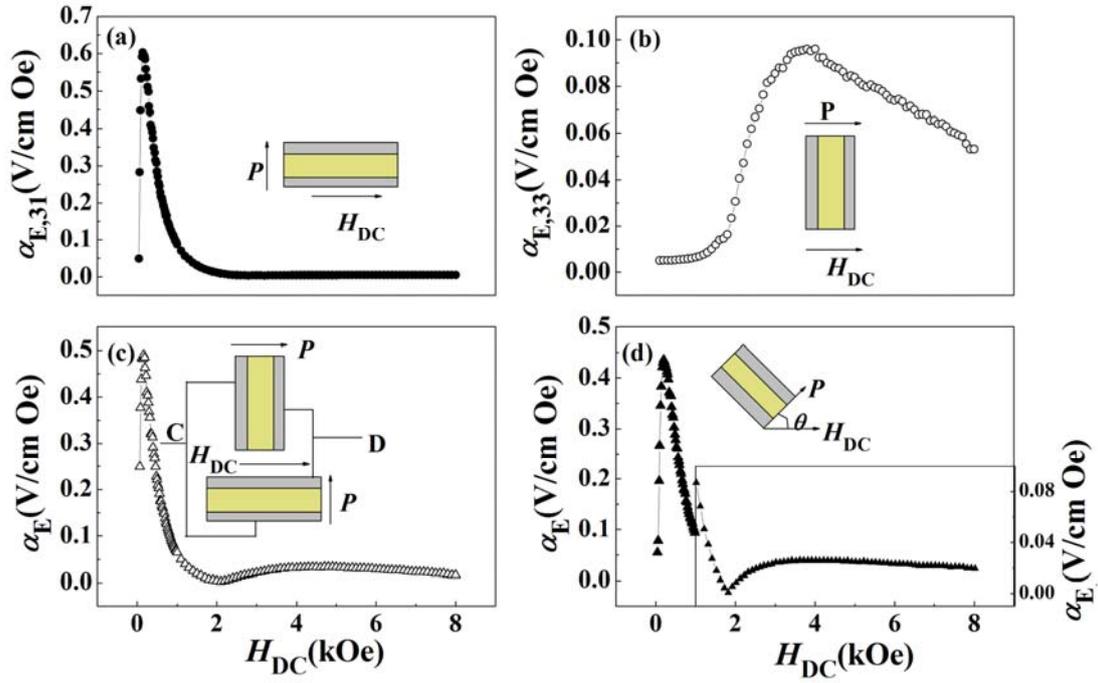

FIG. 3. Dependence of ME voltage coefficient on field $H_{DC}$ at $f$=1 kHz of Ni-PZT-Ni plate composite with dimension of 10×20×0.8 mm$^3$; (a) the plate sample parallel to the field, (b) the plate sample vertical to the field, (c) the parallel connection of two plate samples, (d) the plate composite sloping in the field, $\theta$=45°.

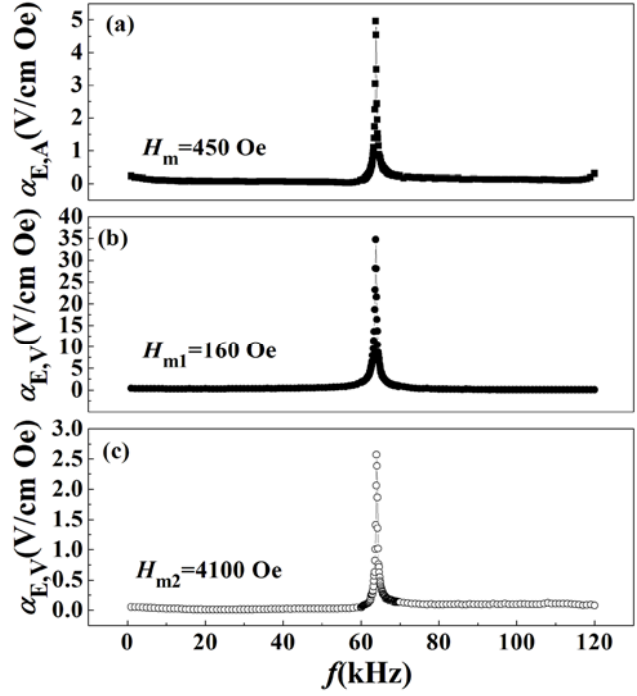

FIG. 4. Frequency dependence of ME voltage coefficient of Ni-PZT-Ni cylindrical layered composite: (a) $\alpha_{E,A}$ at $H_m$=450 Oe, (b) $\alpha_{E,V}$ at $H_{m1}$=160 Oe, (c) $\alpha_{E,V}$ at $H_{m2}$=4100 Oe.

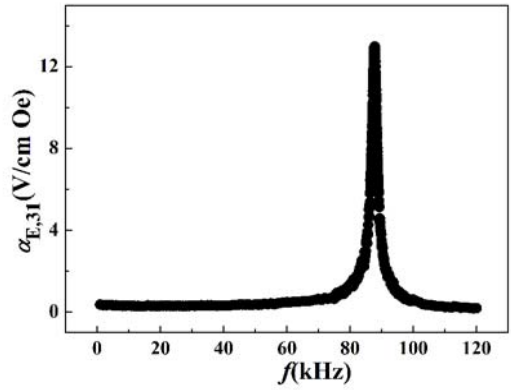

FIG. 5. Frequency dependence of ME voltage coefficient $\alpha_{E,31}$ of Ni-PZT-Ni plate layered composite with dimension of $10\times20\times0.8$ mm$^3$ at $H_{DC}=H_m$. The total thickness of Ni is about 0.8mm.